\documentclass[aps,prd,showpacs,twocolumn,groupedaddress]{revtex4}
\usepackage{graphicx}
\usepackage{amsfonts}
\usepackage{amssymb}
\usepackage{amsmath}

\usepackage{ulem}
\usepackage{color}

\newcommand{\A}{\mathcal{A}}
\begin{document}
\title{Gribov-Zwanziger action
in SU(2) Maximally Abelian Gauge \\ with U(1)$_3$ Landau Gauge }

\author{Shinya~Gongyo}
  \email{gongyo@ruby.scphys.kyoto-u.ac.jp}
  \affiliation{Department of Physics, Graduate School of Science,
  Kyoto University, \\
  Kitashirakawa-oiwake, Sakyo, Kyoto 606-8502, Japan}
  \affiliation{Department of Physics, New York University,
   4 Washington Place, New York, NY 10003, USA}
\author{Hideaki~Iida}
  \email{iida@ruby.scphys.kyoto-u.ac.jp}
  \affiliation{Department of Physics, Graduate School of Science,
  Kyoto University, \\
  Kitashirakawa-oiwake, Sakyo, Kyoto 606-8502, Japan}

\date{\today}
\begin{abstract}
We construct the local Gribov-Zwanziger action for SU(2) Euclidean Yang-Mills theories in the maximally Abelian (MA) gauge with U(1)$_3$ Landau gauge fixing based on the Zwanziger's work in the Landau gauge. By the restriction of the functional integral region to the Gribov region in the MA gauge, we give the nonlocal action. We localize the action with new fields and obtain the action with the shift of the new scalar fields, which has the terms, corresponding to the localized action of the horizon function in the MA gauge. The diagonal gluon propagator in the MA gauge at tree level behaves like the propagator from Gribov-Zwanziger action in the Landau gauge and shows the violation of Kallen-Lehmann representation.
\end{abstract}
\pacs{12.38.Aw, 12.38.Gc, 14.70.Dj}
\maketitle

\section{Introduction}

%
Nowadays, it is thought that the behaviors of the gluon propagator and the ghost propagator are very important for understanding color confinement in Quantum chromodynamics (QCD). In the Landau gauge, confinement is related to the (deep) infrared behavior of the gluon propagator and the ghost propagator from the scenarios proposed by Kugo and Ojima, and Gribov and Zwanziger \cite{KO79,G78,Z89,Z91}.
Gribov and Zwanziger suggested that if the Gribov region is taken into consideration, the gluon propagator is much affected and vanishes at zero momentum.  
From this point of view, the gluon propagator has been much studied by the various methods \cite{AS01,VZ12,P07}.

There are also accurate lattice studies recently which report the infrared behavior of the gluon propagator highly depends on the dimensionality \cite{VZ12,CM07,BIMS09,Ma07,Ma11,DOV10}. In both four and three dimensions, the gluon propagator would be finite at zero momentum \cite{CM07,BIMS09}, which seems to be contradict to the Gribov and Zwanziger scenario mentioned above. In order to solve the challenge, the refined Gribov-Zwanziger action, which introduces the condensates into the original Gribov-Zwanziger action, has been developed \cite{DGSVV08,VZ12}. Compared to these dimensions, the lattice studies in two dimensions show that gluon propagator seems to vanish when the momentum goes to zero \cite{CM07, Ma07}. This indicates that the original Gribov-Zwanziger action is well described in two dimensions at least qualitatively. 

In the maximally abelian (MA) gauge, the gluon propagator has been investigated from the viewpoint of the dual-superconductor picture \cite{GIS12, AS98, BC03, HSA10, Ko11, DGL04}, which is one of confinement scenarios proposed by Nambu, 't Hooft and Mandelstam \cite{N74}. In the dual-superconductor picture in QCD, the abelian degree of freedom leads to the nonperturbative phenomena like confinement (this is called Abelian dominance \cite{tH81,EI82}.) In fact, many studies support that QCD vacuum is the dual superconductor in the MA gauge \cite{KSW87, SY90, BWS91, SNW94, Mi95, Wo95}. 
The Abelian dominance is also supported from the behavior of the gluon propagator. Lattice studies show that the diagonal gluon propagator is larger enhanced than that of the off-diagonal propagator at low momentum in lattice studies \cite{GIS12, BC03}, and thus it indicates the diagonal gluon plays an important role for the infrared physics (especially it is called ``infrared Abelian dominance"). The Schwinger-Dyson study of the gluon propagator also supports the infrared Abelian dominance \cite{HSA10}. 

There are also studies from a perspective of  the Kugo-Ojima (KO) criterion and the Gribov problem. The KO criterion has been generalized to the MA gauge in a very recent paper \cite{MSZA13} and there are also the Gribov copies in the MA gauge, located on the other side of the horizon \cite{Br00,Ba96,Cap06}. 
The gluon propagator in the MA gauge has been also studied on the basis of the original Gribov's work and the horizon term in the MA gauge, corresponding to that in the Landau gauge has been studied \cite{Cap06, G78, Z89}.

On the other hand, the aim of this paper is to construct the local Gribov-Zwanziger action in the SU(2) MA gauge on the basis of Zwanziger's work \cite{Z89}. There are similar studies to generalize the Gribov-Zwanziger action to other gauges \cite{LL11}. In Sec. \ref{2}, we briefly summarize the definition of MA gauge with U(1)$_3$ and the Faddeev-Popov operator in this gauge. In Sec. \ref{3}, we derive the nonlocal action in the MA gauge and localize it with new fields. We consider the vacuum expectation value of the new scalar fields and obtain the final expression of the localized action in Sec. \ref{4} and derive gluon propagators at tree level in Sec. \ref{4.5}. Sec. \ref{5} is devoted to summary and future works.

\section{The Faddeev-Popov operator in SU(2) MA gauge with U(1)$_3$ Landau gauge}
\label{2}
In this section, we review the MA gauge fixing and the Faddeev-Popov operator in SU(2) Euclidean QCD \cite{Cap06}. The local SU(2) MA gauge condition is derived by the local minimization of 
\begin{eqnarray}
R^{\mathrm{MA}}\equiv \sum_{a=1,2}\int d^dx A_\mu ^a (x)A_\mu ^a(x),
\end{eqnarray}
under the gauge transformation. The local form of SU(2) MA gauge is given by
\begin{eqnarray}
\left( D_\mu ^{\rm MA} \right) ^{ab}A_\mu ^b (x)\equiv \left( \partial _\mu \delta^{ab} + \epsilon^{ab}A_\mu ^3 (x)\right) A_\mu ^b (x)= 0, \label{LMAG}
\end{eqnarray}
where the indices $a,b$ are used for off-diagonal components, which run from 1 to 2. 
In this gauge fixing, there remains U(1)$_3$ gauge symmetry, corresponding to the diagonal subgroup of SU(2). We fix the residual symmetry with the Landau gauge,
\begin{eqnarray}
\partial _\mu A _\mu ^3 (x)=0. \label{U1landau}
\end{eqnarray}

The Faddeev-Popov operator in the SU(2) MA gauge, $F$, is given by
\begin{align}
F^{ab}&(x,y)\equiv \left < x,a | F | y,b \right> \nonumber \\ 
&=\left\{-\left(D_\mu ^{\rm MA} \right) ^{ac}\left(D_\mu ^{\rm MA} \right) ^{cb} -\epsilon ^{ac}\epsilon ^{bd} A_\mu ^c A_\mu ^d \right\} \delta (x-y)\nonumber \\
&=\left\{ -\delta ^{ab}\partial ^2 -2\epsilon ^{ab}A_\mu ^3\partial_\mu + \delta^{ab}A_\mu ^3 A_\mu ^3 -\epsilon ^{ac}\epsilon^{bd}A_\mu ^c A_\mu ^d \right\} \nonumber \\
&\times \delta (x-y),  \label{FPdet}
\end{align}
where we used Eq.(\ref{U1landau}). Here, in order to apply the perturbation expansion in the next section, we define the non-interacting part $F_0$ and the interacting part $F_1$ as follows:
\begin{align}
F_0 ^{ab} (x,y)&\equiv -\delta ^{ab}\partial ^2\delta (x-y)  \nonumber  \\
F_1 ^{ab} (x,y)&\equiv \left\{-2\epsilon ^{ab}A_\mu ^3\partial_\mu + \delta^{ab}A_\mu ^3 A_\mu ^3 -\epsilon ^{ac}\epsilon^{bd}A_\mu ^c A_\mu ^d\right\} \nonumber \\
&\times \delta (x-y). 
\end{align}

The Gribov region in the MA gauge is given by the positive definite of the Faddeev-Popov operator,
\begin{eqnarray}
\int d^dx d^dy \psi ^a(x) ^\dagger  F^{ab}(x,y) \psi ^b(y) \geq 0,
\end{eqnarray}
for all vectors $\psi ^a(x)$.
This condition indicates that the Gribov region in the MA gauge is bounded in the off-diagonal direction, while it is not bounded in the diagonal direction \cite{Cap06}. 
\section{The Gribov-Zwanziger action in SU(2) MA gauge with U(1)$_3$ Landau gauge}
In this section, we construct the Gribov-Zwanziger action in the SU(2) MA gauge \cite{Z89}. 
We first express the projected matrix spanned by 
the eigenvectors of the lowest eigenstates of $F_0$ with the perturbation expansion, from which the lowest eigenvalue of $F$ can be obtained. Then we construct Gribov-Zwanziger action 
from the weaker condition of Gribov region mentioned later. 
Localization of Gribov-Zwanziger action is performed in 
the last part of the section.
\subsection{The perturbation expansion at all order in the Faddeev-Popov operator}
\label{3}
To consider the lowest eigenvalue of Faddeev-Popov operator, we perform the perturbation expansion of $F$ around the lowest degenerate eigenvalue of $F_0 $ on the Euclidean volume $V= L^d$. 

The lowest eigenvalue and the eigenstates of $F_0$ are given by
\begin{eqnarray}
F_0\left| k_0,a \right> =k_0^2\left| k_0,a \right> 
\end{eqnarray}
with $k_0 ^2 = \left({2\pi}/{L}\right)^2$. There are 4d eigenstates, corresponding to the wave functions,
\begin{eqnarray}
\left< x_0, b| k_0,a \right> = \delta ^{ab} \left(\frac{1}{L}\right) ^{\frac{d}{2}}e^{ik_0 \cdot x}, \label{wf}
\end{eqnarray}
where $k_0 ^\mu = (0, \ldots ,0, \pm 2\pi/L, 0, \ldots,0)$ are the vectors which have the only one nonzero component, $\pm 2\pi /L$. 

According to Kato's and Bloch's method, the desired eigenvalues of $F$ are those of $\kappa$ given by $4d \times 4d$ matrix \cite{Me65},
\begin{eqnarray}
\kappa \equiv P_0 F U,
\end{eqnarray}
where $P_0$ is the projection operator onto the subspace of the lowest eigenvalue $k_0^2$ of $F_0$,
\begin{eqnarray}
P_0 = \sum_{k_0, a}\left| k_0,a \right> \left< k_0,a \right| ,
\end{eqnarray} 
and $U$ is defined by
\begin{eqnarray}
U&=&\sum _{n=0}^{\infty} U^{(n)}, \\
U^{(n)} &\equiv& \sum_{(n)}{}' S^{p_1}F_1S^{p_2}F_1 \ldots F_1 S^{p_n} F_1 P_0.
\end{eqnarray}
Here,
\begin{eqnarray}
S^p = \left \{ 
\begin{array}{l}
-P_0~~~~~~~~~~~~~~~~~(p=0)  \\
Q_0 \left[\frac{1}{k_0 ^2 - F_0}\right]^p Q_0~~(p\neq 0),
\end{array}
\right.
\end{eqnarray}
where $Q_0 =1-P_0$ and $\sum_{(n)}'$ means the summation over all sets of the non-negative integers $p_1, p_2, \ldots , p_n$ satisfying the conditions,
\begin{eqnarray}
&~&p_1+ p_2 + \ldots + p_m \le m ~ (m = 1,2, \ldots , n-1) \nonumber \\
&~&p_1+ p_2 + \ldots + p_n =n.
\end{eqnarray}
Thus, $\kappa$ is reduced to
\begin{eqnarray}
\kappa &=&k_0 ^2 P_0 + P_0 F_1 \sum _{n=0}^{\infty} U^{(n)} \nonumber \\
&=&k_0 ^2 P_0 + P_0 F_1 \sum _{n=0}^{\infty}\sum_{(n)}{}' S^{p_1}F_1S^{p_2}F_1 \ldots F_1 S^{p_n} F_1 P_0,~~~~~~~ \label{kappa_all}
\end{eqnarray}
where we use $P_0 U = P_0$. In Eq.(\ref{kappa_all}), the terms including $S^0$ are higher order compared to the other terms in the expansion of $1/V$, because the space by $Q_0$ has the additional $V$ factor in the large-volume limit associated with the infinite summation \cite{VZ12}. In the large-volume limit, $\kappa$ is given by
\begin{eqnarray}
\kappa =
k_0 ^2 P_0 + P_0 F_1 \sum _{n=0}^{\infty}S^{1}F_1 S^{1}F_1 \ldots F_1 S^{1} F_1 P_0 \label{kappa_lv}
\end{eqnarray}
Thus, we can rewrite $\kappa$ as a $4d \times 4d$ matrix explicitly,
\begin{eqnarray}
\kappa ^{k_0, k'_0}_{a,b}&\equiv& \left< k_0,a \right| \kappa\left| k'_0,b \right> \nonumber \\
&=&k_0 ^2 \delta_{a,b}\delta_{k_0,k'_0} \nonumber \\
&+&\sum _{n=0}^{\infty} \left< k_0,a \right| F_1
S^{1}F_1  \ldots F_1 S^{1} F_1\left| k'_0,b \right>.~~~~~
\end{eqnarray}
Furthermore, $S^1$ corresponds to $-F_0 ^{-1}$ in the infinite volume limit,
\begin{eqnarray}
S^1 &=& Q_0 \left[\frac{1}{k_0^2 - F_0}\right] Q_0 \nonumber \\
&=& \sum_{|k|>|k_0|,b}\left| k,b \right> \frac{-1}{k^2-k_0 ^2}\left< k,b \right| \nonumber \\
&~&\rightarrow -\sum_{k,b}\left| k,b \right> \frac{1}{k^2}\left< k,b \right|
=-  F_0 ^{-1}, 
\end{eqnarray}
where $k_0 ^2 =(2\pi/L)^2$ is neglected compared to $k^2$.
$\kappa$ is given by
\begin{eqnarray}
\kappa ^{k_0, k'_0}_{a,b} &=&k_0^2 \delta_{a,b}\delta_{k_0,k'_0} 
+\left< k_0,a \right| F_1\left| k'_0,b \right> \nonumber \\
&+&\sum _{n=1}^{\infty} \left( -1 \right) ^n \left< k_0,a \right| F_1
F_0 ^{-1}F_1  \ldots F_1 F_0 ^{-1} F_1\left| k'_0,b \right> \nonumber \\
 &=&k_0^2 \delta_{a,b}\delta_{k_0,k'_0} 
+\left< k_0,a \right| F_1\left| k'_0,b \right>  \nonumber \\
&-&\left< k_0,a \right| F_1 F^{-1}F_1\left| k'_0,b \right>, 
\end{eqnarray}
in the infinite volume limit.

Finally, we give the explicit expression with Eq.(\ref{FPdet}). The second term in $\kappa ^{k_0, k'_0}_{a,b}$ is reduced to 
\begin{align}
\left< k_0,a \right| F_1\left| k'_0,b \right>   
 &=\frac{1}{V}\int d^dx e^{i(k'_0 -k_0)x}\Bigl[-2\epsilon^{ab}A_\mu ^3(x)(ik'_{0 \mu})  \nonumber \\
 &+\delta^{ab}A_\mu ^3(x)A_\mu ^3(x) -\epsilon ^{ac}\epsilon^{bd}A_\mu ^c (x)A_\mu ^d (x)\Bigr],~~~~~~ 
\end{align}
where we used Eq. (\ref{wf}).

Similarly, the third term is rewritten by
\begin{align}
&-\left< k_0,a \right| F_1 F^{-1}F_1\left| k'_0,b \right> \nonumber \\
&=-\frac{1}{V}\int d^dx d^d y e^{i(k'_0 y-k_0x)}\nonumber \\
&\times \Bigl\{-ik_{0\mu}\epsilon^{af} A_\mu ^3(x)
+\delta^{af}A_\mu ^3(x) A_\mu ^3 (x)- \epsilon ^{ac}\epsilon^{fd}A_\mu ^c(x) A_\mu ^d(x)\Bigr\} \nonumber \\
&\times F ^{-1fe} (x,y)\nonumber \\
&\times\Bigr\{ ik'_{0\nu}\epsilon^{eb} A_\nu ^3(y) 
 +\delta^{eb}A_\nu ^3(y) A_\nu ^3 (y)- \epsilon ^{ec}\epsilon^{bd}A_\nu ^c(y) A_\nu ^d(y) \Bigl\} ,
\end{align}
because
\begin{eqnarray}
\left < y, e\right| F_1\left| k'_0, b \right> &=& \frac{1}{\sqrt{V}}( ik'_{0\mu}\epsilon^{eb} A_\mu ^3(y) 
 +\delta^{eb}A_\mu ^3(y) A_\mu ^3 (y)\nonumber \\
&-& \epsilon ^{ec}\epsilon^{bd}A_\mu ^c(y) A_\mu ^d(y))e^{ik_0 y}\nonumber\\
\left < k_0, a\right| F_1\left| x, f \right> &=& \frac{1}{\sqrt{V}}(-ik_{0\mu}\epsilon^{af} A_\mu ^3(x)
+\delta^{af}A_\mu ^3(x) A_\mu ^3 (x)
 \nonumber \\
&-& \epsilon ^{ac}\epsilon^{fd}A_\mu ^c(x) A_\mu ^d(x))e^{-ik_0 x}.
\end{eqnarray}

In this way, we can express $\kappa$ with the inverse of the Faddeev-Popov operator in the MA gauge by resumming the perturbation expansion of $\kappa$.
The lowest eigenvalue of the Faddeev-Popov operator, $\mathcal{E},$ originally belonging to $k_0 ^2$ are obtained from diagonalizing the 4d $\times$ 4d matrix $\kappa$ in principle.

In the next subsection, the Gribov region satisfying $\mathcal{E} > 0$ is replaced by the weak condition, $\mathrm{Tr} \kappa >0$ when we construct the Gribov-Zwanziger action. 
Thus we give $\mathrm{Tr} \kappa $ explicitly:
\begin{eqnarray}
\mathrm{Tr} \kappa = 4dk _0 ^2 + \frac{2d}{V}\kappa _1 + \frac{2k_0^2}{V}\kappa _2 - \frac{2d}{V}\kappa _3,
\end{eqnarray}
where
\begin{eqnarray}
\kappa _1 &=&  \int d^dx\left\{ 2A_\mu ^3(x)A_\mu ^3(x) 
- A_\mu ^a (x)A_\mu ^a (x)\right\}, \nonumber \\
\kappa _2 &=&\int d^dx d^d y \epsilon^{ac} A_\mu ^3 (x) F ^{-1cd} (x,y)
 \epsilon ^{da}A_\mu ^3(y) \nonumber \\
\kappa_3 &=& \int d^dx d^d y \left\{ \delta ^{ac} A_\mu ^3(x)A_\mu ^3(x) - \epsilon ^{ae} \epsilon^{cf}A_\mu ^e(x)A_\mu ^f(x)\right\} \nonumber \\
&\times& F ^{-1cd} (x,y)
 \left\{ \delta ^{da} A_\mu ^3(y)A_\mu ^3(y) - \epsilon ^{de} \epsilon^{af}A_\mu ^e(y)A_\mu ^f(y)\right\} \nonumber \\
 &\equiv &\int d^dx d^d y \A ^{ac}(x) F ^{-1cd} (x,y)\A ^{da}(y) 
\end{eqnarray}
Here, $\kappa _2$ is a higher order correction, because the term has the additional factor $k_0 ^2 = (2\pi /L)^2$, compared to $\kappa_1, \kappa_3$. $\kappa _2$ is discussed in detail in Ref. \cite{Cap06}.

\subsection{Non-local Gribov-Zwanziger action in SU(2) MA gauge with U(1)$_3$ Landau gauge}

The Gribov region is given by restricting the functional region over the gauge field to that of the positive 
Faddeev-Popov operator. Thus, considering the Gribov region, the partition function is expressed by
 \begin{eqnarray}
Z \equiv \int DA D\bar{c}Dc\exp\left[-S_\mathrm{YM}-S_\mathrm{GF}\right] \theta (\mathcal{E}),
\end{eqnarray}
where $S_\mathrm{YM}$ is SU(2) Euclidean Yang-Mills action,
\begin{eqnarray}
S_\mathrm{YM}&=&\frac{1}{4g^2}F^a_{\mu\nu}F^a_{\mu\nu}+\frac{1}{4g^2}F^3_{\mu\nu}F^3_{\mu\nu}\nonumber \\
&=& \frac{1}{4g^2}(\partial_\mu A_\nu^a-\partial_\nu A_\mu^a-\epsilon^{ab}A_\mu^b A_\nu^3+\epsilon ^{ab} A_\mu ^3 A_\nu ^b)^2\nonumber\\
&+&\frac{1}{4g^2}(\partial_\mu A_\nu^3-\partial_\nu A_\mu^3-\epsilon^{ab}A_\mu^a A_\nu^b)^2,
\end{eqnarray}
and $S_\mathrm{GF}$ is the gauge-fixing term in the MA gauge,
\begin{align}
S_\mathrm{GF}&=\frac{1}{2g^2\alpha} \left\{\left( D_\mu ^{\rm MA} \right) ^{ab}A_\mu ^b \right\}^2
+\frac{1}{2g^2\beta}(\partial_\mu A_\mu^3)^2 \nonumber \\
&-\bar c^a F^{ab} c^b -\bar c^3(-\partial^2)c^3-\epsilon^{ab}\bar c^3\partial_\mu (A_\mu^a c^b),
\end{align}
where $c^a$ and $\bar c^a$ are the off-diagonal Faddeev-Popov ghosts and antighosts, 
and $c^3$ and $\bar c^3$ the diagonal Faddeev-Popov ghost and antighost. 
$\alpha$ and $\beta$ are gauge fixing parameters, which are set to be zero 
in the gauge.

The Gribov-Zwanziger action is constructed with the replacement of the condition   $\mathcal{E} >0$ by the weaker condition $\mathrm{Tr} \kappa >0$,  
\begin{eqnarray}
Z_\mathrm{GZ} \equiv \int DA D\bar{c}Dce^{-S_\mathrm{YM}-S_\mathrm{GF}} \theta (\mathrm{Tr} \kappa),
 \end{eqnarray}
where 
\begin{eqnarray}
\theta (\mathrm{Tr} \kappa)&=&\theta ( 4dk_0 ^2 + \frac{2d}{V}\kappa _1+\frac{2k_0 ^2}{V}\kappa_2 - \frac{2d}{V}\kappa _3) \nonumber \\
&=&\theta (2 k_0 ^2V+ \kappa _1 +\frac{k_0^2}{d} \kappa_2- \kappa _3)
\equiv \theta (K).
\end{eqnarray} 

$K$ is exponentiated with the expression, 
\begin{eqnarray}
\theta (K) = \frac{1}{2\pi i} \int _{-\infty}^{\infty} \frac{d\omega}{\omega -i \epsilon} \exp \left(i\omega K \right),
\end{eqnarray}
and the saddle point approximation for the integral,
\begin{eqnarray}
Z_\mathrm{GZ} \simeq \int DA D\bar{c}Dc \exp \left\{-S_\mathrm{YM}-S_\mathrm{GF}+ \lambda K\right\}.
\label{GZaction}
\end{eqnarray}
Here, $\lambda$ is determined from the stationary condition,
\begin{eqnarray}
\frac{\int DA D\bar{c}Dc K\exp \left\{-S_\mathrm{YM}-S_\mathrm{GF}+ \lambda K\right\}}{\int DA D\bar{c}Dc\exp \left\{-S_\mathrm{YM}-S_\mathrm{GF}+ \lambda K\right\}} = \frac{1}{\lambda}, \label{stationary_condition}
\end{eqnarray}
$\lambda >0$ is necessary in order to satisfy the condition (\ref{stationary_condition}) in the Gribov region, $K>0$ \cite{VZ12}.

In this way, we derive the Gribov-Zwanziger action in the SU(2) MA gauge with the U(1)$_3$ Landau gauge,
\begin{align}
Z_\mathrm{GZ} = &\int DA D\bar{c}Dc \nonumber \\
&\exp \left[-S_\mathrm{YM}-S_\mathrm{GF}+ \lambda \left(\kappa _1+\frac{k_0^2}{d}\kappa_2-\kappa _3+2k_0^2 V\right)\right].
\end{align}
Note that in addition to non-local terms, $\kappa _2$ and $\kappa _3$, there are mass terms, $\kappa _1$ due to the nonlinearity of the MA gauge. $\kappa_2$ is localized in Ref. \cite{Cap06} and appendix \ref{ap1}. In the following we show that $\kappa_3$ is also localized and canceled with the mass terms.

\subsection{Localization of Gribov-Zwanziger action in SU(2) MA gauge with U(1)$_3$ Landau gauge}
Next, we introduce a pair of $2\times 2$ complex scalar fields $\left( \phi '^{ab},\bar{\phi} '^{ab}\right)$ in order to localize the non-local Gribov-Zwanziger action in the SU(2) MA gauge. We insert the identity,
\begin{align}
1= \left(\mathrm{det} F\right)^2 \hspace{-0.2cm} \int \hspace{-0.1cm} D \bar{\phi}' D\phi'  \exp \hspace{-0.1cm}\left[ - \hspace{-0.1cm}\int d^dx d^dy \bar{\phi} '^{ba}(x)F^{bc}(x,y) \phi '^{ca}(y) \right]\hspace{-0.1cm} ,
\end{align}
into the partition function $Z_\mathrm{GZ}$ and perform the change of variables,
\begin{align}
\phi '^{ab} (x)&= \phi ^{ab} (x)- \sqrt{\lambda}\int d^dy F^{-1ae}(x,y)\A ^{ec}(y) \epsilon ^{cb} \nonumber \\
&\equiv \phi^{ab}- \sqrt{\lambda}(F^{-1}\A\epsilon )^{ab}, \nonumber \\
\bar{\phi} '^{ab} (x)&= \bar{\phi} ^{ab} (x)- \sqrt{\lambda}\int d^dy \epsilon ^{be} \A ^{ec}(y)F^{-1ca}(y,x) \nonumber \\
&\equiv \bar{\phi} ^{ab}- \sqrt{\lambda}(\epsilon\A F^{-1} )^{ba},
\end{align}

Then the non-local term is canceled,
\begin{align}
\exp&\left[-\lambda\kappa _3\right]=\left(\mathrm{det} F\right)^2 \int D \bar{\phi}D \phi \exp\Biggl[
- \lambda\A ^{ac} \left(F^{-1}\right)^{cd}\A ^{da} \nonumber \\
&- \left[\bar{\phi} ^{ba}- \sqrt{\lambda}(\epsilon \A F^{-1})^{ab}\right]F^{bc}  \left[{\phi} ^{ca}- \sqrt{\lambda}( F^{-1}\A\epsilon )^{ca}\right] \Biggr] \nonumber \\
&= \left(\mathrm{det} F \right) ^2 \int D \bar{\phi}D \phi \nonumber \\
&\hspace{-0.1cm}\exp\Bigg[\hspace{-0.1cm} -\hspace{-0.1cm}\int d^dx d^dy\bar{\phi} ^{ba}F^{bc}\phi ^{ca} 
-\hspace{-0.1cm} \int d^dx  \sqrt{\lambda} \epsilon ^{ba}\A^{bc}\left(\phi ^{ca}-\bar{\phi }^{ca} \right) \hspace{-0.1cm}\Bigg], \label{localkappa3}
\end{align}
where we have used $\A ^{ab} (x)= \A^{ba} (x)$ and omitted the space-time integrals in the first line. Note that the Faddeev-Popov operator is a local operator, and thus the first term in Eq.(\ref{localkappa3}) is localized.

Furthermore, we can localize the determinant term, $\left(\mathrm{det}F\right)^2$, with a pair of ghost fields $\left(\omega^{ab}(x),\bar{\omega}^{ab}(x)\right)$, 
\begin{align}
\left(\mathrm{det}F\right)^2 = \int D\bar{\omega} D\omega \exp \left[ \int d^dx d^dy\bar{\omega}^{ba}(x)F^{bc} (x,y)\omega^{ca} (y)\right].
\end{align}

Similarly, $\kappa_2$ is also localized in appendix \ref{ap1} by introducing a pair of complex vector fields $\left( \psi _\mu^{ab}, \bar{\psi}^{ab} _\mu\right)$ and ghost vector fields $\left( f^{ab} _\mu, \bar{f}^{ab} _\mu\right)$.

Thus, we obtain the local Gribov-Zwanziger action in the SU(2) MA gauge with the U(1)$_3$ Landau gauge,
\begin{align}
Z_\mathrm{GZ} &= \int DA D\bar{c}DcD\bar{\psi}D\psi D\bar{f} Df D \bar{\phi} D \phi D\bar{\omega}  D\omega \nonumber \\
&\exp \Biggl[-S_\mathrm{YM}-S_\mathrm{GF}- \lambda \int d^dx \Big\{ A_\mu ^a A_\mu ^a 
- 2A_\mu ^3 A_\mu ^3\Big\} \nonumber \\
&- \int d^dx d^dy\Big\{\bar{\psi}^{ ba}_\mu F^{bc}\psi^{ca}_\mu -\bar{f}_\mu^{ba}F^{bc} f_\mu^{ca} \Big\}\nonumber \\
&- \int d^dx \sqrt{\lambda \frac{k_0^2}{d}}\epsilon ^{ac} A_\mu ^3  \left(\psi _\mu ^{ca} -\bar{\psi} _\mu ^{ ca} \right) \nonumber \\
&- \int d^dx d^dy \Big\{\bar{\phi} ^{ba}F^{bc}\phi ^{ca} -  \bar{\omega} ^{ba}F^{bc} \omega^{ca} \Big\}\nonumber \\
&- \int d^dx \sqrt{\lambda} \epsilon ^{ba}\A^{bc}\left(\phi ^{ca}-\bar{\phi }^{ca} \right) +2\lambda k_0^2 \int d^dx   \Biggr] 
\end{align}

\section{vacuum expectation values of $\left(\phi,\bar{\phi}\right)$ and the shifted action}
\label{4}
The effective potential of the Gribov-Zwanziger action in the SU(2) MA gauge with the U(1)$_3$ Landau gauge has not the local minimum of $\left<\phi ^{ab}\right> = \left<\bar{\phi} ^{ab}\right>= 0$, but
\begin{align}
\left<\phi^{ab} \right> = -\left<\bar{\phi}^{ab} \right>= \epsilon ^{ab} \sqrt{\lambda},
\end{align}
where we assumed the constant expectation values and used $F^{ac}\phi ^{cb}=F^{ac}\bar{\phi} ^{cb}= \A ^{ac}\phi^{cb}$. 

Then, we shift the variables,
\begin{align}
\phi^{ab}  &\rightarrow  \phi^{ab}  + \epsilon ^{ab} \sqrt{\lambda} \nonumber \\
\bar{\phi}^{ab} & \rightarrow \bar{\phi}^{ab}  - \epsilon ^{ab} \sqrt{\lambda}, 
\end{align}
and obtain the shifted action,
\begin{align}
Z_\mathrm{GZ} &= \int DA D\bar{c}DcD\bar{\psi}D\psi D\bar{f} Df D \bar{\phi} D \phi D\bar{\omega}  D\omega \nonumber \\
&\exp \Biggl[-S_\mathrm{YM}-S_\mathrm{GF}\nonumber \\
&- \int d^dx d^dy\Big\{\bar{\psi}^{ ba}_\mu F^{bc}\psi^{ca}_\mu -\bar{f}_\mu^{ba}F^{bc} f_\mu^{ca} \Big\}\nonumber \\
&- \int d^dx \sqrt{\lambda \frac{k_0^2}{d}}\epsilon ^{ac} A_\mu ^3  \left(\psi _\mu ^{ca} -\bar{\psi} _\mu ^{ ca} \right) \nonumber \\
&- \int d^dx d^dy \Big\{\bar{\phi} ^{ba}F^{bc}\phi ^{ca} -  \bar{\omega} ^{ba}F^{bc} \omega^{ca} \Big\} \nonumber \\
&+2\lambda k_0^2 \int d^dx  \Biggr] ,
\end{align}
where the total derivative terms were neglected. Note that the mass terms including the diagonal part like a tachyonic behavior are canceled and the dependence of $\lambda$ arises only from the well-known horizon term and constant term.

We can easily understand the contributions of $\left( \phi, \bar{\phi} \right)$ and $\left( \omega ,\bar{\omega}\right)$ are trivial, because when they are integrated out, there arise the Faddeev-Popov determinant and the inverse of the Faddeev-Popov determinant. Thus, we obtain the final expression of the Gribov-Zwanziger action in the SU(2) MA gauge with the U(1)$_3$ Landau gauge,
\begin{align}
Z_\mathrm{GZ} &= \int DA D\bar{c}DcD\bar{\psi}D\psi D\bar{f} Df 
\exp \Biggl[-S_\mathrm{YM}-S_\mathrm{GF}\nonumber \\
&- \int d^dx d^dy\Big\{\bar{\psi}^{ ba}_\mu F^{bc}\psi^{ca}_\mu -\bar{f}_\mu^{ba}F^{bc} f_\mu^{ca} \Big\}\nonumber \\
&- \int d^dx \Bigg\{\sqrt{\lambda '}\epsilon ^{ac} A_\mu ^3  \left(\psi _\mu ^{ca} -\bar{\psi} _\mu ^{ ca} \right) +\lambda'  2d \Bigg\}\Biggr] ,
\end{align}
where we defined $\lambda ' \equiv \lambda k_0 ^2/d$.
Note that when we integrate out $\left( \psi, \bar{\psi} \right)$ and $\left( f ,\bar{f}\right)$, the action is just composed of the Yang-Mills term, gauge-fixing term and horizon term as in the Landau gauge. 
The stationary condition is rewritten as
\begin{align}
\frac{\int DA D\bar{c}Dc \left(\kappa_2+2dV\right)\exp \left\{-S_\mathrm{YM}-S_\mathrm{GF}+ \lambda '\kappa_2\right\}}{\int DA D\bar{c}Dc\exp \left\{-S_\mathrm{YM}-S_\mathrm{GF}+ \lambda '\kappa_2\right\}} = \frac{1}{\lambda '}. \label{stationary_condition_final}
\end{align}

\section{gluon propagators at tree level}
\label{4.5}
Now that we have constructed the local Gribov-Zwanziger action in the MA gauge, 
we give the tree-level gluon propagators of the off-diagonal and diagonal parts. The behavior of them in the Gribov-Zwanziger action is 
different for the off-diagonal and diagonal components. 

The propagator of the diagonal gluon in the SU(2) MA gauge with U(1)$_3$ Landau gauge has the form, 
\begin{align}
G_{\mu\nu}^{33}(k)=\frac{k^2}{k^4+4g^2\lambda '}\left[\delta_{\mu\nu}
-\frac{k_\mu k_\nu}{k^2}\right].
\end{align} 
where $\lambda '$ is determined by Eq.(\ref{stationary_condition_final}).   

On the other hand, the propagator of the off-diagonal gluons in the SU(2) MA gauge with the ${\rm U(1)}_3$ Landau gauge is written as 
\begin{align}
G_{\mu\nu}^{ab}(k)=\frac{1}{k^2}\left[\delta_{\mu\nu}
-\frac{k_\mu k_\nu}{k^2}\right]\delta^{ab}.
\end{align}
Note that the diagonal gluon propagator is much affected by the Gribov region and shows the violation of the Kallen-Lehmann representation, while the off-diagonal gluon propagator is not affected by the Gribov region.


\section{Summary and Future works}
\label{5}
We have constructed the local Gribov-Zwanziger action in the SU(2) MA gauge with the U(1)$_3$ Landau gauge fixing. The action has been obtained by restricting the functional integral region of gauge fields and resumming the perturbation of the Faddeev-Popov operator around the lowest (nontrivial) eigenvalue of the unperturbation part under the infinite volume limit. With the condition of positivity of the trace of the Faddeev-Popov operator and the exponentiation of the term, there appear the usual mass terms of the off-diagonal gluons and the anomalous mass term of the diagonal one in the action. These terms show up due to the nonlinearity of the MA gauge condition, Eq. (\ref{LMAG}) and do not show up in the Landau gauge because of the linearity.

In addition to these terms, there appears the nonlocal term with the inverse of the Faddeev-Popov operator which does not correspond to the horizon function in the Landau gauge. Rather, the term corresponding to the horizon function in the Landau gauge, shows up in the next order of inverse volume. The nonlocal term has been also localized by introducing new scalar fields and ghost fields as in the term corresponding to the horizon function \cite{Cap06}. Though the action is localized, the vacuum for the new scalar fields become unstable. Thus we shift the scalar fields to those around the stable vacuum, and then these terms are canceled by the mass terms. In this way, we have obtained the final expression of the local Gribov-Zwanziger action in the SU(2) MA gauge with the U(1)$_3$ Landau gauge, which corresponds to the action including to the Yang-Mills term, gauge-fixing term and horizon term as in Landau gauge.

Furthermore, we have derived the off-diagonal and diagonal gluon propagators at tree level from the Gribov-Zwanziger action in the SU(2) MA gauge with the U(1)$_3$ Landau gauge. The off-diagonal propagator is not affected by the Gribov region, while as for the diagonal part, the propagator seems to show the violation of Kallen-Lehmann representation related to the gluon confinement.

It is also interesting to consider the lower dimensionality. In the Landau gauge, there are also Gribov copies in lower dimensions as in four dimensions \cite{Ma11}.  Especially in two dimensions, the lattice result of the gluon propagator is in agreement with the propagator from the original Gribov-Zwanziger action at least qualitative level \cite{Ma07}.

 Our result in the MA gauge has been obtained in any dimensions. Thus, we can investigate whether the Gribov-Zwanziger scenario is also supported in the MA gauge by comparing our propagators with the two dimensional numerical result in the gauge, though, there are not at present any lattice calculations in two dimensional MA gauge to our knowledge.

\section*{Acknowledgments}
The authors thank Martin Schaden and Daniel Zwanziger for many stimulating discussions and Teiji Kunihiro and Shintaro Karasawa for usuful comments. 
This work is supported in part by the Grant-in-Aid for Scientific Research 
from the JSPS Fellows (No. 24-1458)  
and the Ministry of Education, Culture, Science and Technology 
(MEXT) of Japan (No. 23340067).

\appendix
\section{Localization of the horizon function}
\label{ap1}
In this appendix, we consider the localization of $\kappa_2$, corresponding to the horizon function in the Landau gauge \cite{Cap06}. 
$\kappa_2$ term in the Gribov-Zwanziger action Eq.(\ref{GZaction}) is localized by inserting the identity,
\begin{align}
1&= \left(\mathrm{det} F\right)^{2d} \int D\bar{\psi}'_\mu D\psi '_\mu \nonumber\\
&~~~~~~~~~~~~~~~\exp \left[ - \int d^dx d^dy \bar{\psi}'^{ ba}_\mu F^{bc}(x,y) \psi'^{ca}_\mu \right] ,
\end{align}
where we introduced a pair of complex bosonic vector fields, $( \psi^{ba}_\mu , \bar{\psi}^{\dagger ba}_\mu )$. The non-local term is canceled by the change of variables,
\begin{align}
\psi^{ba}_\mu (x)= \psi'^{ba}_\mu (x)- \sqrt{\lambda \frac{k_0 ^2}{d}}\int d^dy F^{-1bc}(x,y) \epsilon ^{ca}A_\mu ^3 (y),
\end{align}
and $(\mathrm{det} F)^{2d}$ is also localized by introducing a pair of fermionic vector fields $(f_\mu^{ab} , \bar{f_\mu}^{ab})$,
\begin{align}
\left(\mathrm{det} F\right)^{2d} = \int D\bar{f} Df \exp \left[ \int d^dx d^dy \bar{f}_\mu^{ba}F^{bc} f_\mu^{ca} \right].
\end{align}
In this way, $\kappa _2$ term is localized \cite{VZ12,Cap06},
\begin{align}
&\exp \left[\lambda\frac{k_0 ^2}{d}\kappa_2 \right]= \int D\bar{f} Df D\bar{\psi}_\mu D\psi \nonumber \\
& \exp \Bigg[- \int d^dx d^dy\Bigg\{\bar{\psi}^{ ba}_\mu F^{bc}\psi^{ca}_\mu -\bar{f}_\mu^{ba}F^{bc} f_\mu^{ca} \Bigg\}\nonumber \\
&- \int d^dx \sqrt{\lambda \frac{k_0^2}{d}}\epsilon ^{ac} A_\mu ^3  \left(\psi _\mu ^{ca} -\bar{\psi} _\mu ^{ ca} \right)\Bigg].
\end{align}
The Fadeev-Popov operator is a local operator and thus we obtain the local form of all terms in the Gribov-Zwanziger action of the SU(2) MA gauge, Eq. (\ref{GZaction}).

\end{document}